# System Synchronization Based On Complex Frequency

Lan Tang, *Member, IEEE*, Yusen Wei, *student Member,IEEE*
and Peidong Li, *student Member, IEEE*

***Abstract*—The increasing penetration of renewable energy leads to a continuous reduction in system inertia, for which conventional synchronization criteria based solely on frequency consistency can no longer accurately capture the coupled dynamics of frequency and voltage during transients. To address this issue, this paper employs the concept of complex frequency and develops an analysis framework that integrates theory, indices, and simulation for assessing synchronization stability in low inertia power systems. First, the basic concepts and mathematical formulation of complex frequency and complex frequency synchronization are introduced. Then, dynamic criteria for local and global complex synchronization are established, upon which a complex inertia index is proposed. This index unifies the supporting role of traditional frequency inertia and the voltage support capability associated with voltage inertia, enabling quantitative evaluation of the strength of coordinated frequency–voltage support and disturbance rejection within a region. Finally, transient simulations on a modified IEEE 9 bus system are carried out to validate the proposed method. The results show that the method can clearly reveal the synchronization relationships between subnetworks and the overall system, providing a useful theoretical reference for stability analysis and control strategy design in low inertia power systems.**

## I. INTRODUCTION

In conventional AC power systems, synchronization is typically defined as the uniformity of rotor angle variation rates among all synchronous generating units in the network [1].Studies on synchronization can essentially be regarded as a class of transient stability analysis, whose primary objective is to ensure that system voltage and frequency remain stable following disturbances [2].In classical power systems dominated by synchronous generators, transmission lines are generally strongly inductive $(X \gg R)$ .As a result, practical engineering control strategies often treat active power–frequency (P/f) control and reactive power–voltage (Q/V) control as approximately decoupled, handling voltage regulation and frequency regulation in separate loops [3]. However, such decoupled treatment neglects the intrinsic coupling between the electromagnetic and mechanical states inside a synchronous generator, namely the interaction between rotor mechanical dynamics and excitation system dynamics. When the system is subjected to large disturbances or requires rapid regulation, this interaction can be significantly amplified, leading to strong coupling and interference between the voltage control loop and the frequency control loops. The resulting additional frequency and voltage oscillations may adversely affect the transient stability of the overall system [4]-[8].

With the large scale integration of high penetration renewable resources, synchronous generators and their rotational inertia are being extensively replaced by power electronic interfaced devices. As a result, the equivalent system inertia is significantly reduced, the frequency dynamics become much faster, and the system exhibits markedly increased sensitivity to active power disturbances [9]-[11]. Under these conditions, even small to medium scale disturbances may trigger noticeable discrepancies in inter area frequency responses and transient desynchronization between local subnetworks and the bulk system, thereby increasing the risk of system separation, load shedding, and even regional blackouts. If synchronization is still judged solely by the uniformity of electrical angle variation rates, it becomes difficult to promptly identify and accurately characterize these emerging instability mechanisms.Recent studies have shown that, in low inertia power systems, the coupling between active power frequency and reactive power voltage is significantly strengthened, and the time scales of voltage and frequency regulation tend to converge. Local voltage disturbances can propagate through control loops and network power flows and, in turn, feed back adversely on frequency stability [12],[13].Meanwhile, most generation and load units employing AC regulation are equipped with fast voltage and current control loops, rendering the node voltage magnitude and its rate of change increasingly critical to disturbance propagation and inter area interactions. In recent years, a complex frequency analysis framework has been proposed and applied to transient studies of power systems, in which the real part of the complex frequency represents the normalized rate of change of node voltage magnitude, while the imaginary part corresponds to the conventional system frequency deviation, thereby capturing the coupled voltage frequency dynamics within a unified mathematical description[14].

Building on this framework, the concepts of complex frequency synchronization and associated stability criteria have been further developed to analyze multivariable synchronization behavior in converter dominated systems, revealing the respective advantages and drawbacks of voltage frequency interactions among different regions[15],[16].Therefore, introducing the notion of complex frequency synchronization, which characterizes the coordinated evolution of local voltage magnitude and phase angle speed, is of substantial theoretical and practical significance for identifying steady state instability risks induced by local voltage frequency coupling under high renewable penetration, and for enhancing the accuracy of dynamic stability assessment and control strategy design in low inertia power systems.

## II. Complex Frequency Synchronization Principle

### A. Complex Frequency

The conventional complex frequency is defined as the complex variable in the Laplace transform,$S = \sigma + j\omega$,where the real part $\sigma$ is the damping factor and the imaginary part $\omega$ is the conventional instantaneous angular frequency[17].However, with the large scale integration of inverter based resources (IBRs), the coupling between system frequency and voltage has become increasingly pronounced, and this traditional definition of complex frequency can no longer accurately capture the dynamic characteristics of electrical quantities during transients.

To address this issue, F. Milano proposed a new definition of complex frequency $\eta = \varepsilon + j\omega$[14], in which the real part $\varepsilon$ represents the normalized rate of change of the voltage magnitude, while the imaginary part $\omega$ remains the conventional instantaneous angular frequency. In this way, voltage and frequency can be described within a unified framework. Fig. 1 compares the complex frequency responses of a system dominated by synchronous generators and a low inertia system dominated by power electronic converters. In the conventional power system, the disturbance response of the real part of the complex frequency exhibits a relatively small amplitude and a time scale that is clearly shorter than that of frequency oscillations of electrical quantities, so that the transient dynamics of voltage and frequency are relatively decoupled. In contrast, in the low inertia power system, the real and imaginary parts of the complex frequency display almost identical evolution, and their time scales become markedly aligned, reflecting a much stronger coupling between voltage and frequency. This comparison further illustrates the necessity of introducing complex frequency based unified descriptions and synchronization criteria under low inertia conditions.

A brief mathematical derivation of the new complex frequency concept is given below. The voltage at an arbitrary bus $k$ in the network $(k \in \mathbb{N})$ can be expressed as

$$\bar{u}_k = u_d + ju_q = ue^{j\theta} \tag{1}$$

Rewriting (1) as $\bar{\vartheta} = e^{lnv+j\theta}$,where $\vartheta$ denotes the complex angle[14],and differentiating (1) with respect to time yields

$$\dot{\bar{u}} = \left(\frac{\dot{v}}{v} + j\dot{\theta}\right)\bar{u} \tag{2}$$

Letting $U = \ln(v)$,(2) can be simplified as

$$\dot{\bar{u}} = \left(\dot{U} + j\dot{\theta}\right)\bar{u} \tag{3}$$

Define the complex frequency as, $\eta = \left(\dot{U} + j\dot{\theta}\right) = \varepsilon + j\omega$ [14].

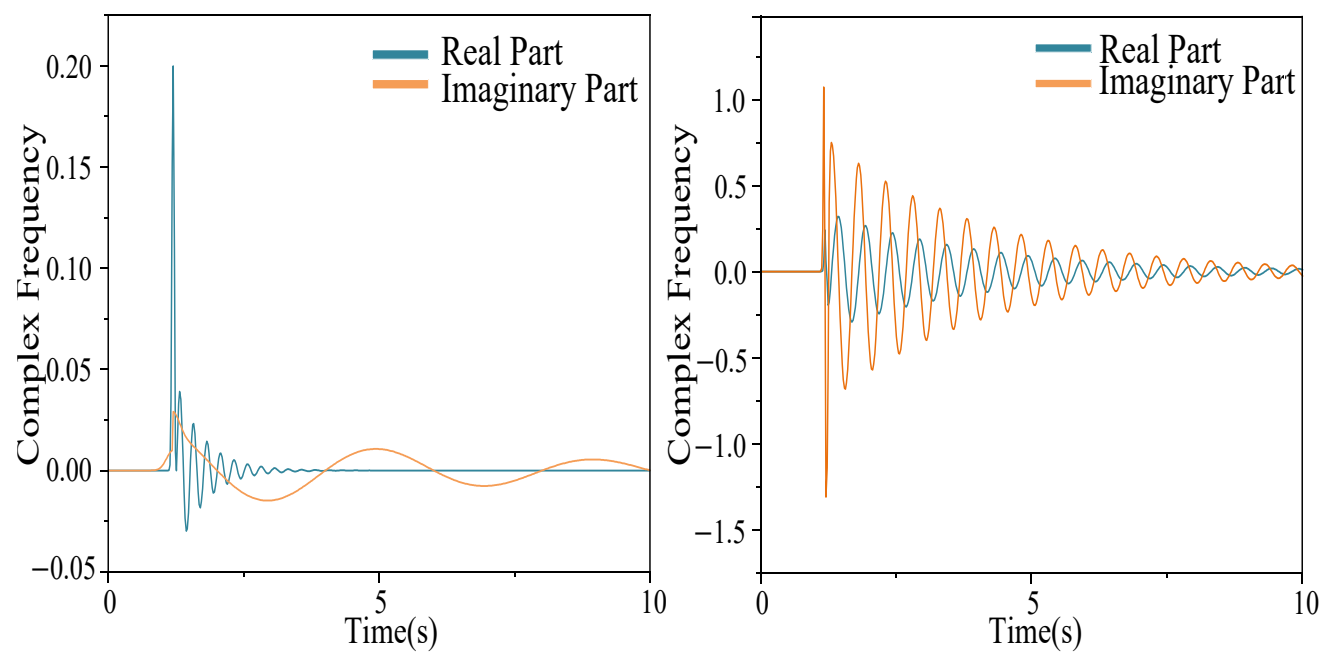

Figure 1. Variation in the Voltage Frequency Time Scale Coupling Relationship

*B. Fundamental Principle of Complex Frequency Synchronization*

Complex frequency synchronization refers to the dynamic process in which, after a disturbance or control action has settled, the complex frequency $\bar{\eta} = \varepsilon + j\omega$ at every node of the system converges to the same constant value over the entire network[18]. Its rigorous mathematical definition is

$$\lim_{t\to\infty} \bar{\eta}_k(t) = \lambda, \quad \forall k \in \mathcal{N} \tag{4}$$

where, when the complex frequencies of all nodes in the region converge to a constant value $\lambda$, the system is said to achieve complex frequency synchronization.

Starting from the node voltage dynamics, a sufficient condition for such convergence is derived below.

From (2), the complex frequency at node $k$can be written as, $\eta_k(t) = \frac{\dot{v}_k(t)}{v_k(t)}$.

where $v_k(t)$ denotes the voltage signal at node $k$ ,and $v_k(t) = e^{\vartheta}$ ,with $\vartheta$ being the complex phase angle in (1).

Solving the above DAE by separation of variables and integrating from the initial time $t_0$ to $t$ yields the expression of $v_k(t)$ as

$$v_k(t) = v_k(t_0) e^{\int_{t_0}^{t} \eta_k(s) ds} \tag{5}$$

When $t \to \infty$, the integral term of $\eta_k(s)$ in (4) can be decomposed into a limit value and a deviation term, namely

$$\int_{t0}^{t} \eta_k(s) ds = \lambda t - \lambda t_0 + C_k(t) \tag{6}$$

Where,$C_k(t) = \int_{t0}^{t} (\eta_k(S) - \lambda) ds$

Equation (5) indicates that the deviation term is absolutely integrable, which ensures that its integral evolves as a convergent dynamic process as $t \to \infty$,and its limit is finite.

From basic mathematical relations, if the absolute integral of a function is finite, then the integral of the function itself is also finite and convergent. Hence,

$$C_k(\infty) = \lim_{t\to\infty} C_k(t) = \int_{t0}^{\infty} (\eta_k(s) - \lambda) ds \tag{7}$$

Substituting (6) into the target expression $e^{-\lambda t} v_k(t)$ .As $t \to \infty$, the limit of (6) becomes

$$\phi_k = v_k(t_0) \cdot e^{(C_k(\infty) - \lambda t_0)} \tag{8}$$

Therefore, one obtains $e^{-\lambda t} v_k(t) \to \phi_k(t \to \infty)$This shows that convergence of the complex frequency together with absolute integrability of its deviation constitutes a sufficient condition for complex frequency synchronization of the system.Complex frequency synchronization not only requires conventional frequency uniformity, but also imposes a stricter requirement that voltage dynamics exhibit consistent behavior in both the time domain and the direction of magnitude variation. This provides a more comprehensive index framework for transient and voltage stability assessment in high penetration, low inertia power systems[19].

*C. Scope of Synchronization Criteria*

The concept of complex frequency synchronization provides a generalized and complementary extension of the conventional notion of synchronization at the theoretical level. When the system voltage magnitude is essentially constant（$\rho = 0$）,this concept reduces to classical synchronization in the sense of uniform angular frequency[14].From a control perspective, complex frequency synchronization represents a new synchronization paradigm: it treats the normalized rate of change of voltage magnitude $\rho$ and the angular frequency $\omega$ as joint synchronization objectives, thereby requiring new measurement and control architectures (such as coupled $[\omega, \rho]$ based droop schemes) and significantly improving the robustness of frequency signals in the early stages following a disturbance. To clarify the practical applicability of complex frequency synchronization, its scope of use is further delineated below.

1) Application Scenarios of Complex Frequency Synchronization

In the following typical scenarios, focusing solely on conventional rotor angle synchronization is clearly insufficient, and complex frequency synchronization is required to capture the full system dynamics.

a) **Systems with high penetration of inverter based resources (GFM/GFL).**In such systems, rotational stability is directly determined by electrical and control dynamics, and the inertial response of conventional synchronous machines is effectively replaced by virtual inertia. At the same time, voltage side dynamics have a pronounced impact on the transient response. When the penetration level of grid following (GFL) converters exceeds about 60% voltage dynamics or excitation related dynamics tend to emerge or deteriorate before frequency synchronization issues become evident[20]. Hence, it is necessary to monitor both $\rho$ and $\omega$ in order to detect instability risks at an early stage.

b) **Operating conditions involving phase locked loops (PLLs)**.such as grid connection or mode switching [21]. When responding to faults, switching events, or other disturbances, PLLs are prone to abrupt frequency changes or oscillations. Complex frequency can provide a smoother and more reliable frequency indicator, and the inclusion of the voltage magnitude rate $\rho$ helps mitigate the signal distortion associated with using $\omega$ alone.

c) **Distribution networks with low $X/R$ ratios, or systems whose electrical dynamics have time scales comparable to those of the control loops.** When line resistance cannot be neglected, the coupling between active power/frequency and reactive power/voltage is significantly strengthened (with approximate relationships such as ($\dot{q}' \approx G'\omega$, $\dot{p}'' \approx G''\rho$),so traditional observations based solely on $\omega$ will overlook the influence of voltage dynamics on frequency behavior [14].In addition, in low inertia systems, the time scales of line dynamics and AC filter control dynamics can be comparable, making the conventional assumption of neglecting network dynamics no longer valid.

d) **Islanded or weakly coupled operating areas.** During transients, significant discrepancies may arise among the nodal frequencies $\omega$ [22], so relying solely on a single measurement point or on the system center of inertia (COI) frequency cannot accurately assess the synchronization status. In this case, employing a regionally consistent complex frequency variable $\eta$ can enhance the overall coordination of the system and improve the robustness of stability and synchronization criteria.

e) **Control scenarios with coupled frequency and voltage oscillations.** The complex frequency synchronization framework indicates that, even in lossy networks, frequency oscillations can be suppressed through coordinated active and reactive power control. Simulation studies further show that frequency regulation based on the combined use of $\rho$, $\omega$, and $\eta$ achieves better performance than schemes relying solely on active power $p$[23].In essence, this requires treating $\omega$and $\rho$as joint control objectives.

2) Application Scenarios of Conventional Synchronization

Although complex frequency synchronization provides a more general framework for analyzing the stability of low inertia power systems, this does not imply that traditional synchronization criteria based solely on the angular frequency $\omega$ have become invalid. Therefore, after clarifying the application scenarios of complex frequency synchronization, it is still necessary to systematically review and delineate the conditions under which conventional synchronization remains applicable. When the system dynamics satisfy, or are close to satisfying, the following assumptions, monitoring $\omega$alone is sufficient to accurately characterize the synchronization status.

a) **Systems in which synchronous generators (SGs) dominate or the voltage stiffness is relatively high**[24].When the system is dominated by SGs and the automatic voltage regulator (AVR) together with the power system stabilizer (PSS) can effectively maintain the bus voltage magnitude nearly constant, one has $\rho \approx 0$. In this situation, system dynamics are mainly reflected in the frequency channel, and complex frequency synchronization naturally degenerates into conventional angular frequency synchronization. Existing work[21] also indicates that, as long as the system is able to preserve a firm voltage profile, stability issues arise primarily on the frequency side; only when the share of converter interfaced generation (CIG) keeps increasing and the firm voltage condition is weakened do both state variables become necessary.

b) **Systems dominated by grid forming converters (GFMs) that do not rely on phase locked loops (PLLs) for synchronization.** In GFM based

architectures providing system wide voltage and frequency support, the frequency $\omega$at the point of common coupling is prescribed by internal control rather than obtained via PLL tracking. This yields a frequency signal that is largely immune to grid disturbances. If, in addition, the AC system experiences only small variations in voltage magnitude, $\omega$can be regarded as a reliable indicator of the synchronization state[25].

c) **Quasi steady state operation or small disturb ance conditions.** Reference[26] emphasizes that, under the quasi steady state assumption, frequ ency differences among system nodes are extre mely small; in this regime, $\rho$is typically close t o zero, and $|\omega|$ is usually much larger than $|\rho|$. In such cases, neglecting the influence of $\rho$ and focusing solely on $\omega$ constitutes a reaso nable and sufficiently accurate engineering appr oximation.

## III. Design of Robustness Indices for System Synchronization

To further quantify the dynamic behavior of a system under disturbances, a set of synchronization performance indices is developed on the basis of the proposed complex frequency synchronization criterion. Conventional same frequency criteria focus mainly on frequency dynamics and are therefore inadequate for capturing the coupled voltage frequency characteristics of low inertia systems. In this context, two key indices are introduced in this paper. The first is the oscillation decay rate, which is used to evaluate the damping of transient dynamics during the synchronization process. The second is the complex inertia, which provides an overall measure of the system's capability to support both voltage and frequency. Taken together, these two indices offer complementary tools for synchronization assessment and control design, from the viewpoints of dynamic evolution and steady state equivalence, respectively[27].

### *A. Oscillation Decay Rate*

The oscillation decay rate is an index characteri zing the dynamic process by which a system restore s synchronization after a disturbance; physically, it p rovides a quantitative description of the local dampi ng characteristics of the system and is therefore cru cial for identifying weakly damped components in fr equency oscillations[28].

Within the complex frequency framework, the real and imaginary parts of the complex frequency at node $k$ can, after a disturbance, be approximated by exponential decay models:

$$|\varepsilon_k(t) - \overline{\varepsilon}_k| \approx A_{\varepsilon,k} e^{-\sigma_{\varepsilon,k}} \tag{9}$$

$$|\omega_k(t) - \overline{\varepsilon}_k| \approx A_{\omega,k} e^{-\sigma_{\omega,k}} \tag{10}$$

Where,$\sigma_{\varepsilon,k}$ and $\sigma_{\omega,k}$ are defined as the oscillation decay rates of the voltage component and frequency component at node $k$, respectively. Larger values of these parameters indicate faster decay of the corresponding oscillations, stronger damping, and higher robustness of the synchronization process. Conversely, if the decay rates are significantly small, the associated node may be subject to sustained oscillations, potentially threatening the overall stability of the system.

At the subnetwork or system wide level, the overall dynamic performance of a region can be evaluated by analyzing the statistical distribution of nodal decay rates. If multiple nodes within a given subnetwork exhibit uniformly low decay rates, that region can be identified as a critical area with insufficient damping and should be prioritized for controller parameter optimization or the installation of additional damping devices.

### *B. Complex Inertia*

In conventional AC power systems, the inertia constant $M$characterizes the capability of synchronous generators to withstand sudden changes in rotor speed and determines the rate of frequency decline following an active power imbalance. Since electromagnetic transients are much faster than electromechanical ones, engineering analysis typically adopts the approximate decoupling between $P/f$and $Q/V$channels[29].However, with the large scale integration of high penetration renewable resources, this separation of time scales is significantly weakened: voltage and frequency evolve on comparable time scales, and their mutual dynamic responses are of the same order of magnitude. As a result, relying solely on the mechanical inertia $M$ is no longer sufficient to fully describe the transient response characteristics of the system.

To jointly capture these two types of dynamics within a unified framework, a voltage inertia $H_v$ is introduced on the basis of $M$ to measure the inertial support capability of the system against voltage variations, and the two are further combined into the complex inertia $\zeta = H_{v,k} + jM_{\omega,k}$A theoretical derivation of the complex inertia concept is given below.

Consider the Norton equivalent at the target node, whose total terminal admittance is denoted as[24].

$$Y_{eq}(j\omega) = Y_{th}(j\omega) + Y_{device}(j\omega) \tag{11}$$

where $Y_h$ is the equivalent admittance of the external network as seen from the node (including lines, shunt compensators, transformers, etc.), and $Y_{device}$ is the admittance of the interfaced device (grid connected converters, synchronous machines, etc.) together with its voltage/reactive power control loops, all referred to the terminal. The reactive power is written as

$$Q = V^2 B_{eq}(\omega),\ B_{eq} = \Im(Y_{eq}) \tag{12}$$

According to Foster's theorem[30], under a given terminal voltage, the reactive energy at the port can be expressed as

$$W_{vis}(V_i, \omega) = \frac{1}{4} V_i^2 \frac{d}{d\omega}(\omega B_{eq}(\omega)) \tag{13}$$

Equation (13) unifies the physical shunt capacitance and the virtual capacitance produced by control into the equivalent susceptance $B_{eq}$. Accordingly, the voltage inertia constant and voltage stiffness at node $i$ are defined as

$$H_{v,i} \triangleq \frac{V_{i0}^2}{4S_{base}} \frac{d}{d\omega}(\omega B_{eq,i}(\omega)) \tag{14}$$

In (14), $K_{v,i}$is the voltage stiffness proposed in[24], used to describe the voltage support capability of node $i$, whereas $H_{v,i}$quantifies the node's resistance to the rate of change of voltage (RoCoV).

The voltage support capability of a node can b e characterized by its ability to inject capacitive rea ctive power, whose physical essence is a capacitive support effect[31].

Therefore, the frequency dependent characteristic of $B_{eq}$ is mapped to explicit equivalentcapacitance parameters that simultaneously reflect static support and dynamic regulation capabilities, defined as

$$C_{eq}^{stat} \triangleq \frac{B_{eq}(\omega_0)}{\omega_0},\ C_{eq}^{dyn} \triangleq \frac{dB_{eq}}{d\omega} \tag{15}$$

Here, the static equivalent capacitance $C_{eq}^{\mathrm{stat}}$ represents the effective capacitive energy storage capability, while the dynamic capacitance $C_{eq}^{\mathrm{dyn}}$quantifies the slope of the equivalent susceptance $B_{eq}$ with respect to frequency, capturing the dynamic response of the port to external frequency disturbances. $B_{eq}$can be approximated as

$$B_{eq}(\omega) \approx \omega_0 C_{eq}^{stat} + (\omega - \omega_0) C_{eq}^{dyn} \tag{16}$$

When (16) satisfies $\frac{d}{d\omega}(\omega B_{eq}) \approx C_{eq}$, the expression of the voltage inertia constant degenerates to

$$H_{v,i} = \frac{V^2 C_{eq}}{S_{base}} \tag{17}$$

The mechanical swing equation can be written as

$$\frac{2H}{\omega_s}\frac{d\omega}{dt} = P_m - P_e \tag{18}$$

Introducing the real part of the complex frequency into (17) gives

$$\varepsilon(t) \triangleq \frac{\dot{V}(t)}{V(t)} = \Re\{\overline{\omega}(t)\} \tag{19}$$

Substituting (13) into the energy balance relation yields

$$\dot{W}_{vis} \approx \frac{1}{2} V^2 \frac{d}{d\omega}[\omega B_{eq}]|_{\omega_0} \varepsilon = \Delta Q \tag{20}$$

Combining (17) and (20), the following equivalent relationship between the voltage inertia constant and the real part of the complex frequency is obtained

$$\frac{V_0^2}{2S_{base}} C_{eq} \cdot \varepsilon \triangleq H_v \varepsilon = \Delta Q \tag{21}$$

In summary, two parallel power inertia relations are derived

$$\begin{aligned} M\dot{\omega} &= \Delta P \\ H_v \varepsilon &= \Delta Q \end{aligned} \tag{22}$$

Define, $\zeta = H_v\varepsilon + jM\dot{\omega} = H_{v,k} + jM_{\omega,k}$,where $\zeta$ is termed the complex inertia.

The concept of complex inertia characterizes th e dynamic disturbance rejection strength of a local r egion in both the voltage and frequency dimensions. By jointly computing and comparing, for each nod e, the voltage inertia contribution$H_{v,k}$and the frequen cy inertia contribution $M_{\omega,k}$, one can quantitatively evaluate the voltage support and frequency support capabilities of each subnetwork. This approach effect ively identifies regional networks with insufficient p ost disturbance voltage support or weak damping of frequency oscillations, thereby providing theoretical guidance for targeted reactive power compensation a nd optimal placement of energy storage resources.

## IV. CASE STUDIES

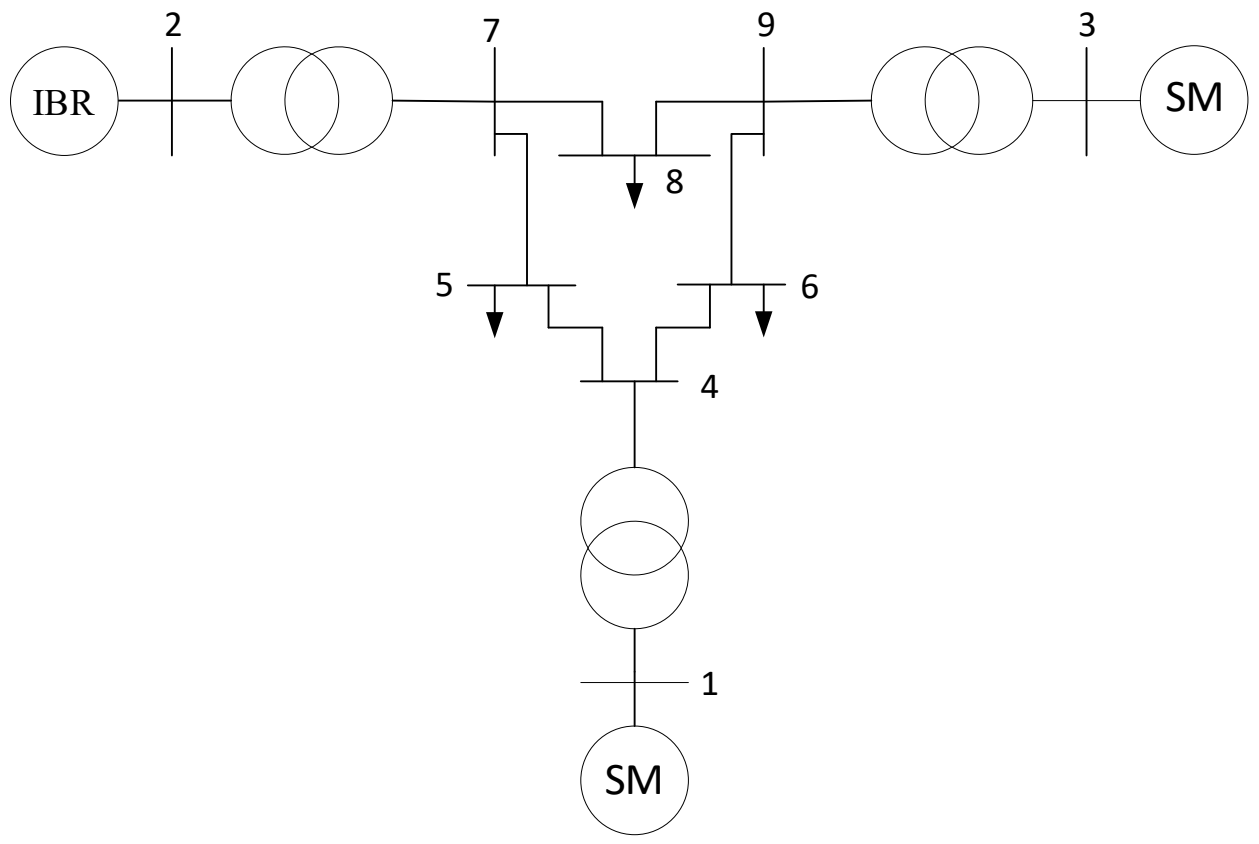

Figure 2. Modified WSCC 9-bus system

To verify the effectiveness of the proposed complex frequency synchronization criterion and its associated indices, a modified WSCC-9 bus system is established using the open-source tool Andes[32] and transient simulations are carried out for analysis[33]. In this model, each synchronous generator is equipped with a complete excitation system and governor, while the synchronous generator at Bus 2 is replaced by a photovoltaic generation unit. According to the electrical coupling between each generator and its neighboring load buses, the system buses are partitioned into three subnetworks: S1 (Bus2, Bus7, Bus5), S2 (Bus1, Bus4, Bus6), and S3 (Bus3, Bus8, Bus9).

Section 4.1 investigates the local synchronizatio n characteristics of the system through time domain simulations. Based on the aforementioned synchroniz ation criterion, the analysis starts from the converge nce behavior of individual buses and is then extend ed to the synchronization within each subnetwork. O n this basis, Section 4.2 evaluates the global synchr onization performance of the overall system from th e perspective of subnetwork synchronization, and fur ther examines the oscillation decay rate and comple x inertia synchronization indices.

### *A. Local Synchronization Experiment*

To analyze the synchronization characteristics of the system, the following disturbance scenario is considered: at $t = 2$ s, the load connected to Bus 6 is tripped. The complex frequency at each bus is measured using the *BusFreq* module.

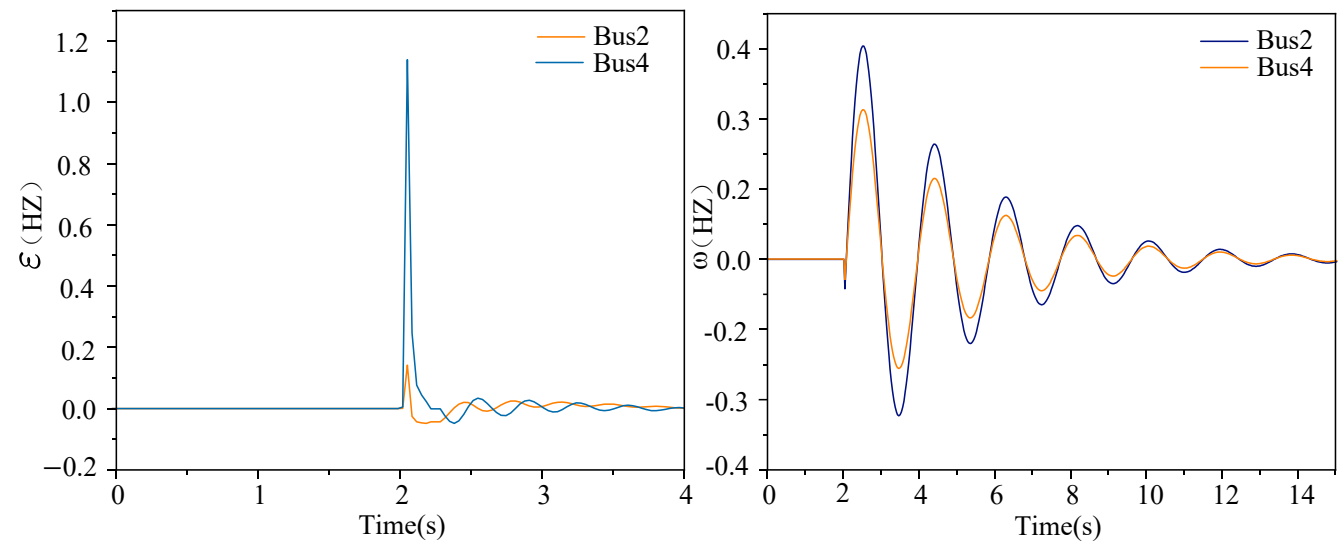

Figure 3. Complex Frequency curves for Bus 2 and Bus 4

Figure 3 compares the post disturbance synchro nization dynamics of Buses 2 and 4. The left panel plots the real part of the complex frequency, $\varepsilon$(Hz). When the disturbance occurs at $t = 2$ s, both buses exhibit a pronounced overshoot in $\varepsilon$, and their dec ay trajectories and convergence speeds are clearly di fferent. The right panel shows the conventional angu lar frequency $\omega$. During the transient, the two curve s gradually decay and almost coincide. Under traditi onal synchronization criteria that only examine whet her frequency or rotor angle converges to a commo n steady state value, the system would therefore be judged as having returned to a stable operating con dition. However, Fig. 3 reveals that the real parts at the two buses display evident overshoot at the dist urbance instant, and their decay paths and converge nce rates differ markedly. This indicates that the rat e of change behavior of the voltage magnitude does not achieve uniform convergence, and that there ex ists dynamic inconsistency that is obscured when on e looks only at frequency.

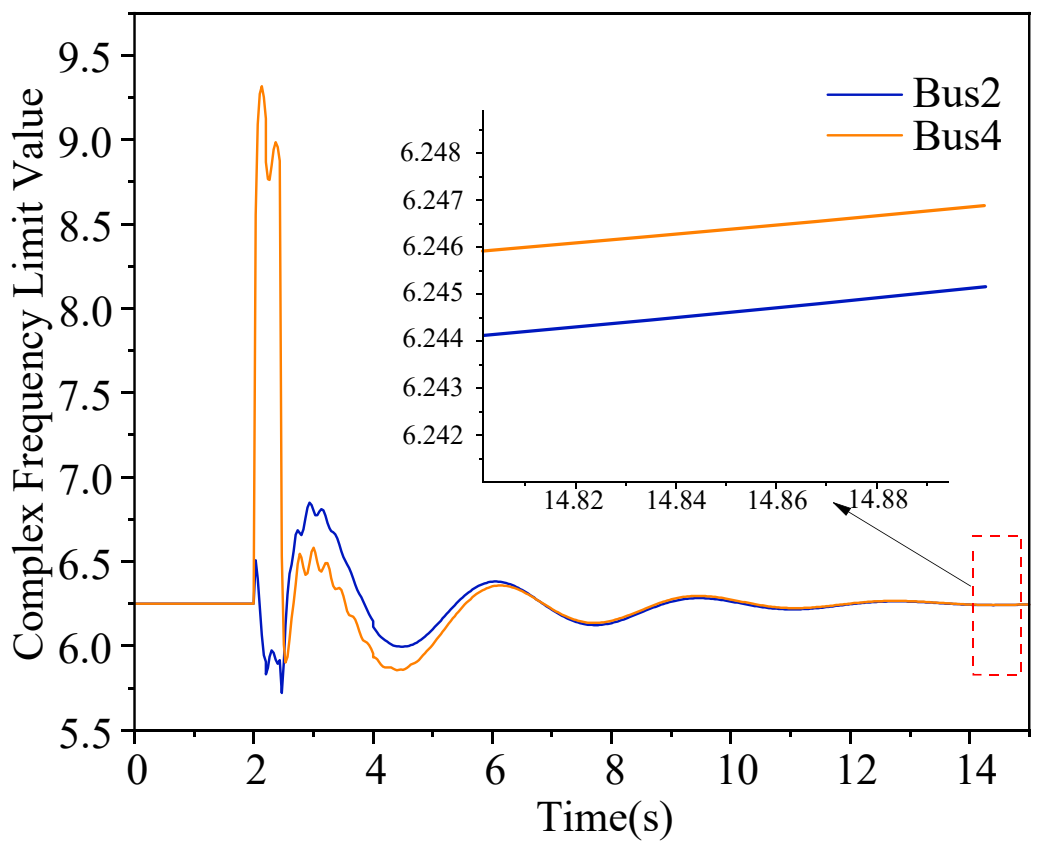

Figure 4. complex synchronization curves for Bus 2 and Bus 4

Figure 4 further demonstrates that the complex synchronization criterion requires not only that the bus frequencies approach the same value in steady state, but also that the voltage frequency dynamic responses remain coordinated throughout the entire transient process. The results show that, although the system appears to exhibit a synchronous trend in terms of frequency, the analysis from the complex frequency perspective indicates that the system is actually in a nonsynchronized state. This phenomenon is consistent with the definition of complex frequency synchronization given in this paper, and it reveals the potential risk associated with apparent synchronization but internal desynchronization: differences in voltage recovery rates among buses will aggravate reactive power oscillations, weaken the overall damping of the system, and make it difficult for local oscillations to decay synchronously across the network [34]. More critically, if different buses respond differently to the same disturbance, protection devices may misoperate or lose coordination, thereby jeopardizing the secure and stable operation of the system [35].

To investigate the synchronization tendency of each subnetwork under disturbance, time domain simulations are carried out for the above contingency. Figure 5 shows the post disturbance complex frequency response of subnetwork S1.

Immediately after the disturbance at $t = 2$ s, all three trajectories exhibit a pronounced overshoot and then enter a decaying process; although their initial magnitudes differ, the overall oscillation patterns remain consistent. Around $t \approx 14$ s, the three curves converge to the same steady state value, indicating that, within subnetwork S1, the nodal rate of change of voltage magnitude and the angular frequency achieve coordinated responses in both dimensions, thereby satisfying the strict definition of complex frequency synchronization.

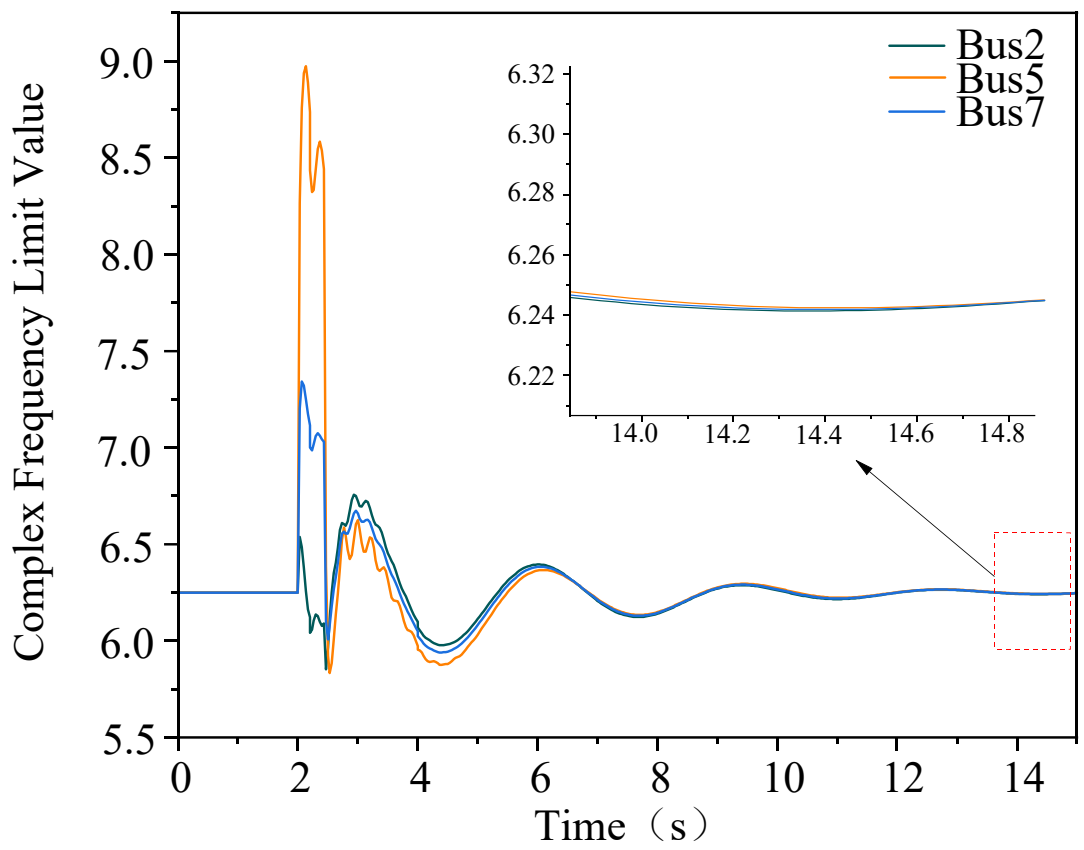

Figure 5. Complex Synchronization Curve for Subnetwork S1

Figure 6 presents the dynamic response of the complex frequency limit values at buses 3, 8, and 9 in subnetwork S2. The responses at buses 3 and 8 show larger amplitudes with more pronounced oscillations, whereas the trajectory at bus 9 is relatively smooth. As the transient progresses, the system gradually regains synchronization: although the three curves differ in oscillation amplitude and phase, their oscillatory and decaying trends are essentially consistent. The magnified inset clearly shows that, for $t \geq 12$ s, all three trajectories converge to the same steady state value.

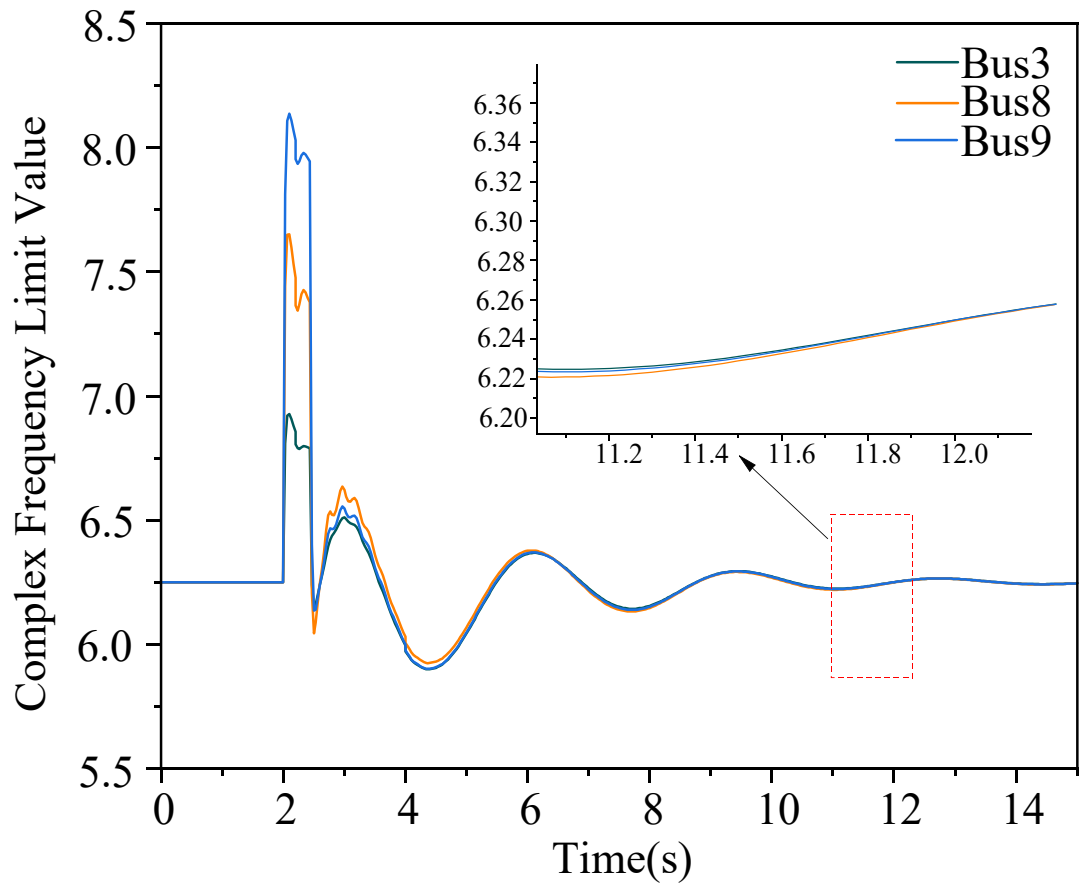

Figure 6. Complex Synchronization Curve for Subnetwork S2

From the comparative analysis of the complex frequency synchronization processes in subnetworks

S1 and S2, significant differences can be observed in synchronization speed and dynamic convergence characteristics. In the time domain, subnetwork S2 exhibits superior synchronization performance: it reaches a stable state roughly 6 safter the disturbance, and the oscillation amplitudes of its nodal trajectories are smaller with faster decay, reflecting a larger decay rate, stronger damping, and better coordinated dynamic response. By contrast, the synchronization process in subnetwork S1 lasts for about 12 s; its internal nodes show evident overshoot and sustained oscillations after the disturbance, with a relatively low decay rate, indicating weaker local dynamic coupling. Although both subnetworks eventually achieve complex frequency synchronization, S2 clearly outperforms S1 in terms of synchronization speed and dynamic quality. This difference further confirms the effectiveness of complex frequency as an index for evaluating subnetwork synchronization performance, and provides theoretical support for identifying weak regions in the system and for optimizing local controller parameters.

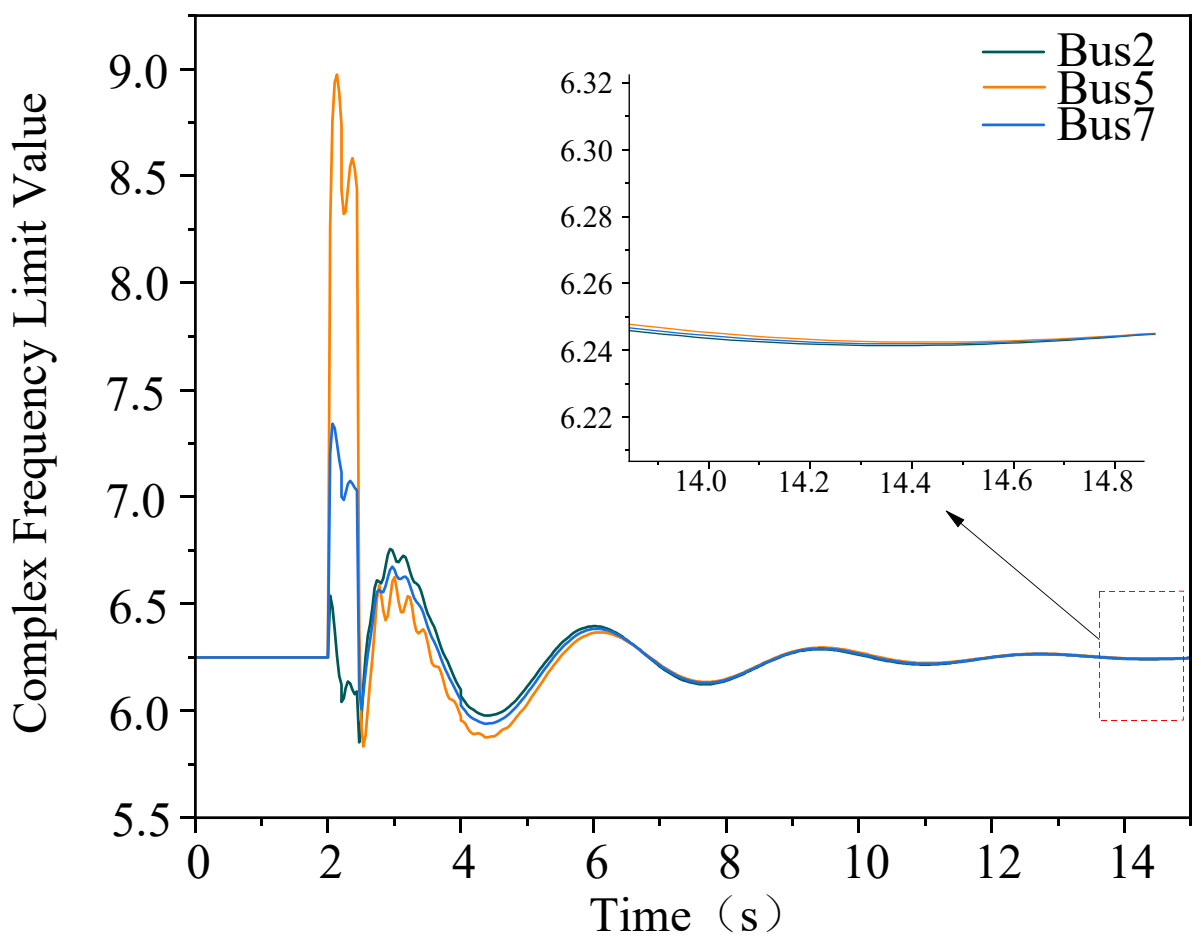

Figure 7. Complex Synchronization Curve for Subnetwork S3

Figure 7 shows the complex frequency synchronization response of subnetwork S3 under the disturbance. As seen in the figure, all three trajectories exhibit a pronounced overshoot immediately after the disturbance and then enter a decaying oscillatory stage, among which the transient magnitude at Bus 5 is the largest, indicating that this node bears a relatively stronger impact. As the transient evolves, the three trajectories display highly consistent oscillatory decay patterns and remain closely synchronized in their dynamic evolution. The enlarged inset indicates that, for $t \geq 12$ s, all curves converge to the same steady-state value, and the oscillation decay during the transient is well coordinated, with no local loss-of-step phenomena. This response characteristic demonstrates that S3 achieves strict complex-frequency synchronization: its internal nodes exhibit good dynamic coordination in both the voltage rate of change and angular frequency dimensions. Compared with S1 and S2, the synchronization process of S3 is smoother, reflecting stronger regional damping and better dynamic coherence.

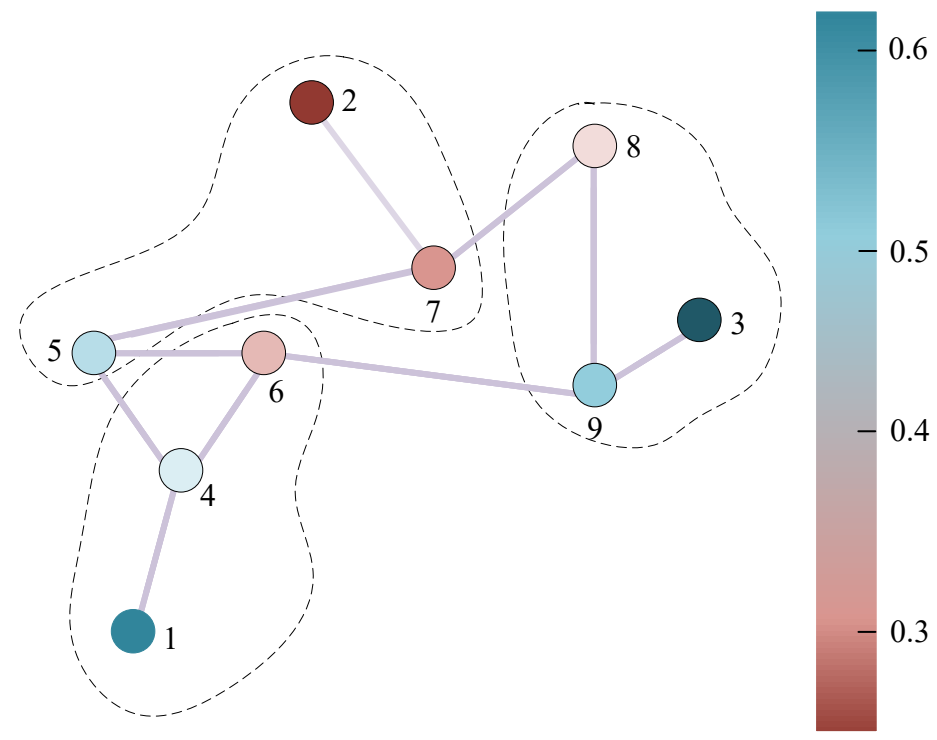

Figure 8. Heatmap of Oscillation Attenuation Rate Distribution for 9 Nodes

Figure 8 depicts the spatial heat map distribution of oscillation decay rates for all nodes in the 9 bus system. Regions in cool colors correspond to higher decay rates and therefore stronger damping performance, whereas regions in warm colors indicate lower decay rates and slower oscillation attenuation. From the nodal perspective, Buses 2 and 7 clearly appear in warm tones, and Bus 5 lies between warm and neutral tones, implying that these three nodes dissipate post disturbance oscillation energy more slowly overall. In contrast, Buses 3, 8, and 9 are mainly shown in cool or slightly cool colors, and their oscillation decay rates are markedly higher, revealing stronger local damping capability. Extending the analysis of oscillation decay rates from individual nodes to regional areas thus helps identify subregions with weak damping, and provides a theoretical basis for subsequent optimization of damping related control parameters and the siting of additional damping devices.

*B. Global Synchronization Experiment*

In Section 4.1, the complex frequency synchron ization trajectories of the three subnetworks were pr esented separately, and it was shown that their post disturbance convergence speeds and damping levels differ significantly. However, comparisons restricted t o the internal behavior of each subnetwork are not sufficient to capture their integrated impact on the o verall dynamic coordination of the system. To assess the global complex synchronization process among the three subnetworks under the same disturbance sc enario (load shedding at Bus 6 at $t = 2$ s), it is nec essary to introduce a *global network synchronization curve*. Specifically, the complex frequency limit valu es of S1, S2, and S3 are aggregated at the subnetw ork level so as to obtain trajectories that reflect the evolution of global coherence.

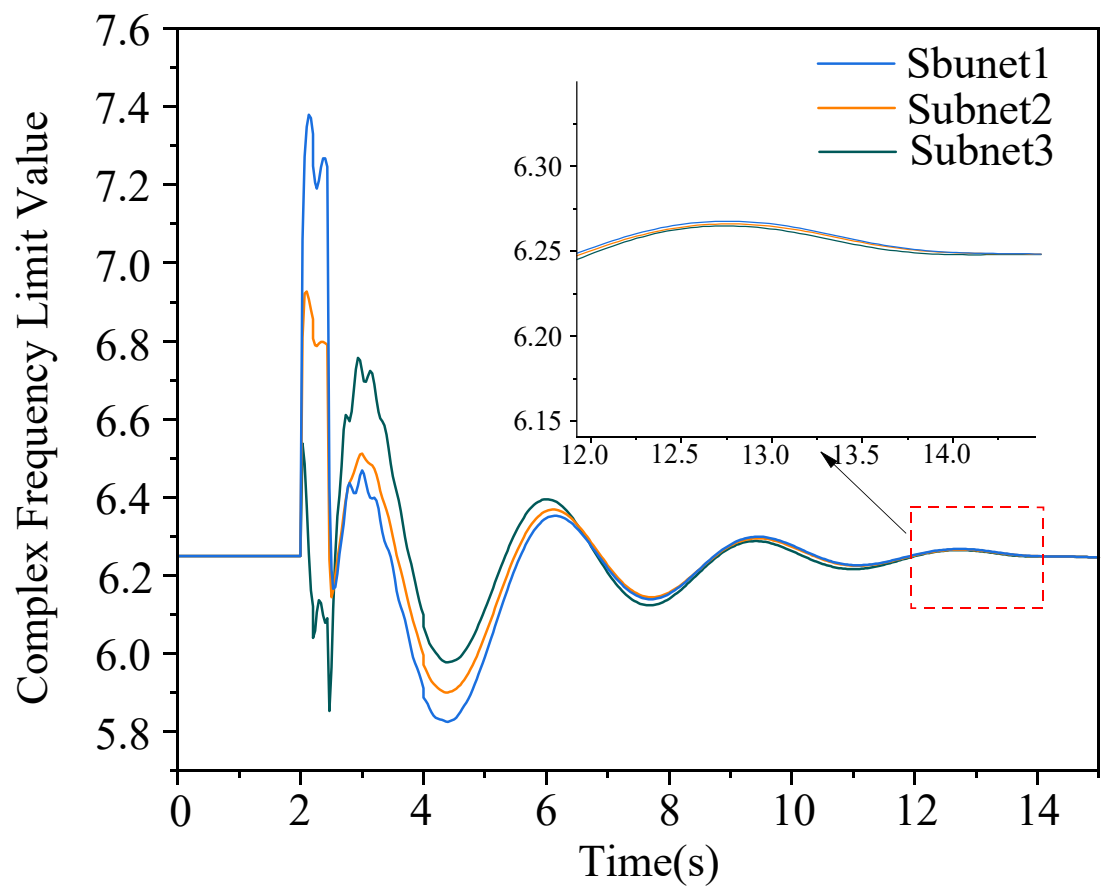

Figure 9. Global Network Synchronization Curve

As shown in Fig. 9, the three curves almost coincide before the disturbance, indicating that the entire network is operating at a common steady state point. Once the load at Bus 6 is removed at $t = 2$ s, the complex frequency limit values of the three subnetworks diverge rapidly. Among them, S1 exhibits the largest response magnitude: both its initial overshoot and the subsequent reverse swing over a certain period are significantly larger than those of S2 and S3, and its oscillation decay process lasts longer, with the trajectory remaining visibly separated from the other two curves during approximately$6-8$ s. By contrast, the synchronization curves of S2 and S3 change more smoothly, with smaller oscillation amplitudes and an earlier entry into the decay phase. The magnified window in the interval $12-14$ s further shows that the trajectory of S1 still lies slightly above those of S2 and S3, implying that its complex frequency convergence remains relatively slow even near steady state.

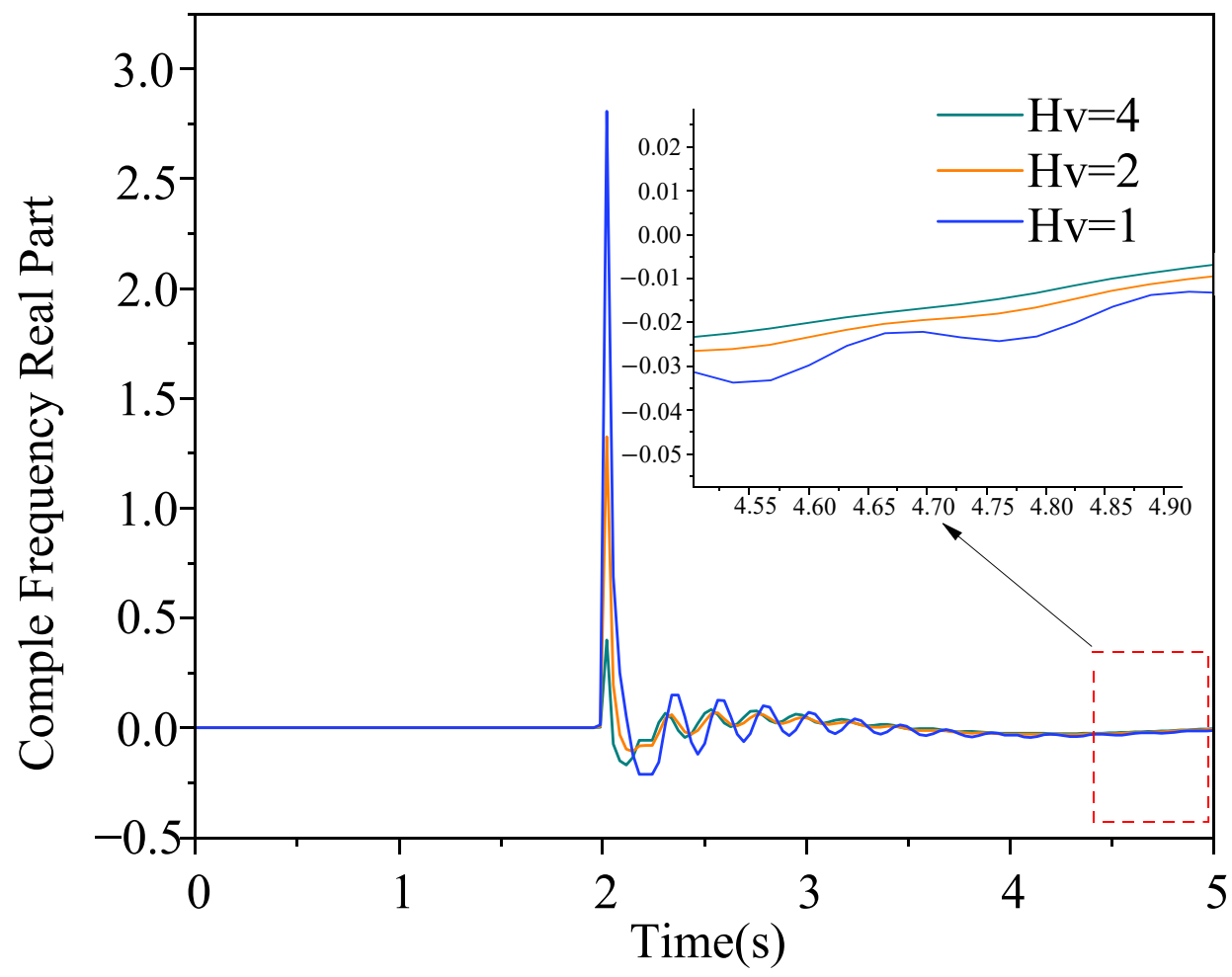

Figure 10. Rate of Change in the Real Part of the Complex Frequency at Different $H_v$ Values

Figure 10 compares the proposed complex inertia index using the real part of the complex frequency. It shows the transient responses of the real part of the complex frequency at Bus 4 under different voltage inertia parameters $H_v$. Before the disturbance, all three curves remain flat around zero, indicating that the system operates at a steady state with zero rate of change of voltage magnitude. After the load at $t = 2$ sis shed, the three trajectories all exhibit a pronounced positive spike followed by a decaying oscillation. Among them, the case with $H_v = 1$ has the largest peak and the strongest oscillation; when $H_v = 2$, both the peak value and oscillation amplitude are reduced; when $H_v = 4$, the initial spike is clearly suppressed and the subsequent oscillations decay rapidly, so that the overall response is characterized by smaller peaks and smoother fluctuations. In the enlarged steady state transition region it can be seen that, for larger $H_v$, the real part of the complex frequency after the disturbance decays closer to the zero axis and varies more smoothly. This indicates that increasing the voltage side inertia effectively mitigates abrupt changes and oscillations in the voltage magnitude rate, thereby enhancing the system's dynamic voltage support capability under disturbances.

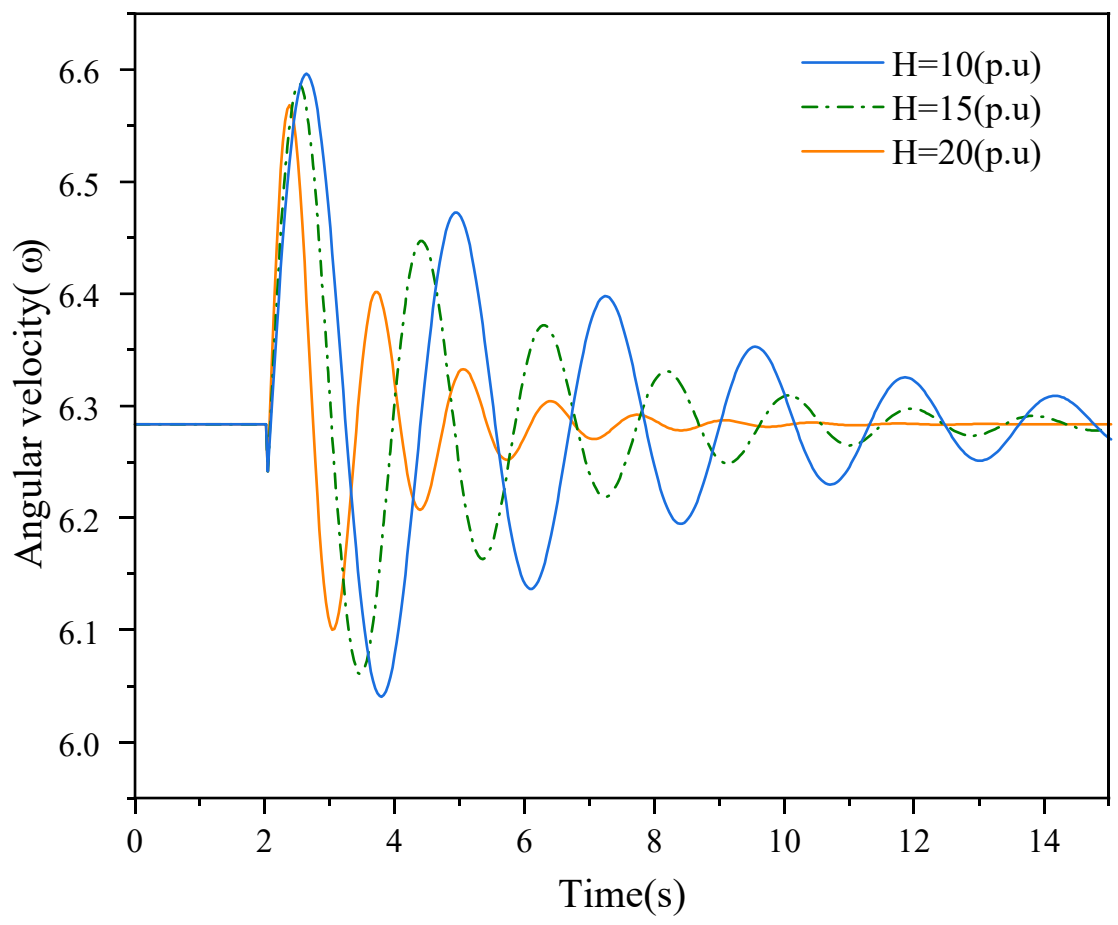

Figure 11. System Frequency Variation under Different Inertia Parameters $H$

Figure 11 presents the rotor speed (frequency) response under different rotational inertia parameters $H$, again for the case of a load step at $t = 2$ s. When $H = 10$, the initial overshoot of the frequency deviation and the subsequent oscillation amplitude are the largest, and the time required for the frequency to settle around the new equilibrium is also relatively long. As $H$ is increased to 15 and 20, the initial frequency deviation progressively decreases, the oscillation peaks and troughs are clearly suppressed, and for $H = 20$ the frequency enters a decaying oscillatory regime more rapidly and converges to its steady state value. The similarity in waveform characteristics between the two figures shows that the voltage side inertia parameter $H_v$ and the conventional mechanical inertia $H$ play analogous inertial buffering roles in their respective channels: the former improves voltage dynamic performance by suppressing abrupt changes and oscillations in the real part of the complex frequency, while the latter enhances frequency stability by reducing frequency deviations and accelerating the decay of frequency oscillations. Together, they constitute the complex inertia framework defined in this paper, providing a solid basis for unified assessment of voltage frequency coupled dynamics.

## V. Conclusion

Building on the theory of complex frequency, this paper proposes a complex frequency based synchroniza tion criterion suitable for partitioned power systems, w hich can uniformly describe the synchronization status of the system at three levels: node, subnetwork, and o verall network. Transient simulations on a modified W SCC 9 bus system, together with comparisons against traditional synchronization criteria that rely solely on fr equency consistency, demonstrate that the proposed met hod can still distinguish the numerical differences betw een the real and imaginary parts of the complex frequ ency even when the system appears frequency synchro nized. In doing so, it clearly reveals local and inter su bnetwork desynchronization phenomena and accurately captures the adverse impact of delayed local complex synchronization on the global synchronization process.

On this basis, two quantitative indices are further introduced: the oscillation decay rate and the complex inertia. The oscillation decay rate characterizes the damping level of nodes and subnetworks and the dissipation rate of oscillation energy; its spatial distribution helps identify poorly damped regions and provides guidance for regional damping enhancement and control resource allocation. The complex inertia index extends the conventional notion of rotational inertia by introducing the concept of voltage inertia, thereby unifying the dynamic response characteristics on the frequency and voltage sides. It enables a joint assessment of the disturbance rejection capability of local areas in both dimensions. Simulation results show that increasing voltage inertia not only suppresses abrupt changes and oscillations in the real part of the complex frequency, but also yields response dynamics that are consistent with the effect of traditional inertia on frequency behavior, thus validating complex inertia as an effective composite indicator of system dynamic strength.

Future work can proceed along two directions. First, the complex synchronization criterion and complex inertia index will be systematically embedded into existing stability assessment frameworks to develop unified methods that simultaneously cover voltage stability and angle stability, and their applicability will be validated under multiple operating scenarios. Second, leveraging the spatial distribution characteristics of oscillation decay rate and complex inertia, we will investigate control and optimization strategies for distribution level renewable dominated power systems, including virtual inertia and voltage inertia allocation, coordinated active reactive power control, and partitioned operation and protection setting design under complex synchronization constraints, so as to provide new theoretical support and engineering approaches for the dynamic security and stable operation of power systems.